\documentclass[12pt,twoside]{article}
\usepackage{a4wide,epsf}
\newcommand{\vskipone}{\vspace{14.5pt plus 2pt minus 2pt}}
\newcommand{\vskiptwo}{\vspace{29pt plus 4pt minus 3pt}}

\begin{document}
\vglue 20mm

{\parindent=0mm {\large QCD factorization in $\gamma^*\gamma\to\pi\pi$
and $\gamma^*N\to\pi\pi N$}%
\renewcommand{\thefootnote}{}\footnote{\hspace{-21pt} To be published in
the proceedings of the conference INPC'98, Paris, France, August
1998}\vskipone

M. Diehl$^{a}$, T. Gousset$^{b}$, B. Pire$^{c}$ and
O.V. Terayev$^{d}$\vskipone

$^a$ DESY, 22603 Hamburg, Germany\vskipone

$^b$ SUBATECH, B.P. 20722, 44307 Nantes, France\vskipone

$^c$ CPhT, Ecole Polytechnique, 91128 Palaiseau,
France\vskipone 

$^d$ Bogoliubov Laboratory, JINR, 141980 Dubna, Russia}
\vskiptwo

Exclusive two-pion production near threshold in the collision of a
highly virtual photon with a real one offers a new possibility to
unravel the partonic content of hadrons. We explain the dynamics of
this regime, i.e.\ the separation of the amplitude in terms of
partonic diagrams computable in QCD perturbation theory and
generalized distribution amplitudes, which are nonperturbative
functions describing the exclusive transition of a parton pair into
two hadrons. The same quantities also appear in $\gamma^*N\to\pi\pi N$
in the kinematical regime where the momentum transfer of the nucleon
and the invariant mass of the $\pi\pi$ pair are both small; this
allows to extend the QCD analysis of exclusive $\rho$
electroproduction outside the resonance region of the two produced
pions.
\vskiptwo

\noindent{\bf 1. FACTORIZATION}\vskipone

Factorization is the property that allows the description of a
hadronic process in terms of quarks and gluons. It means the
separation of a perturbatively computable subprocess, dominated by a
{\em hard} momentum scale, from {\em soft} nonperturbative hadron
matrix elements. A well-known example of this is inclusive deep
inelastic scattering $\gamma^\ast p \to X$, where the cross section is
given by the imaginary part of the forward amplitude $\gamma^\ast p
\to \gamma^\ast p$ via the optical theorem. In the Bjorken limit this
amplitude can be calculated as a perturbative parton-photon scattering
times a parton distribution in the proton. Parton distributions,
which can be extracted from the structure function $F_2 (x_B,Q^2)$,
are universal objects, appearing in other hard processes like lepton
pair production $p+p\to\mu^+\mu^-+X$ at large invariant mass, and the
production of heavy gauge bosons $W$ or $Z$.

Another case of factorization occurs in hard exclusive processes,
e.g.\ meson or proton electromagnetic form factors, or exclusive
electroproduction of vector or pseudoscalar mesons. For instance, the
pion form factor may be written as a convolution of the form
\vskipone

\noindent $\displaystyle
F_\pi(Q^2)=\int_0^1 dz\int_0^1 dz'\,
\varphi(z;\mu^2)\,H(z,z',Q^2;\mu^2)\,\varphi(z';\mu^2),
$\vskipone

\noindent where $H(z,z',Q^2;\mu^2)$ is the perturbatively calculable
hard scattering amplitude of a virtual photon on a $q\bar{q}$ pair at
factorization scale $\mu^2$, and $\varphi(z;\mu^2)$ is the pion
distribution amplitude.

In the following we present a further instance of factorization, which
opens a new domain of investigation for the quark structure of
hadrons. For further details we refer to~[1].
\vskiptwo

\noindent{\bf 2. THE PROCESS $\gamma^*\gamma\to\pi\pi$}
\vskipone

To leading order in $\alpha_S$ the scattering amplitude of the
reaction $\gamma^*(q) + \gamma (q') \to \pi(p) + \pi(p')$
at large $Q^2=-q^2$ and small $W^2=(p+p')^2$ can be represented as
(see Figure~1)\vskipone

\noindent $\displaystyle
T^{\mu\nu}= {1\over 2} (v^\mu v'^\nu\!+v'^\mu v^\nu\!-g^{\mu\nu})
\sum_q e_q^2\int_0^1\!\!dz\,{2z-1\over z(1-z)}\,
\Phi_q(z,\zeta,W^2;Q^2).
$\vskipone

\noindent Here we use lightlike vectors, $v$ and $v'$, defining a
``plus'' and a ``minus'' direction, which are related to the photon
momenta by\vskipone 

\noindent $\displaystyle q = (v-v')\,Q/\sqrt{2}\quad$ and
$\quad\displaystyle q'=v'\,(Q^2+W^2)/(\sqrt{2}\,Q)$.
\vskipone

\noindent $z$ and $\zeta$ are the ``plus'' momentum fraction carried
by the quark and by the $\pi^+$, respectively:\vskipone

\noindent $\displaystyle
z={k\cdot v'\over (p+p')\cdot v'},\quad
\zeta={p\cdot v'\over (p+p')\cdot v'}.
$\vskipone

\noindent The generalized distribution amplitude (GDA) $\Phi_q$ is
defined through the soft matrix element that describes the $q\bar q
\to \pi\pi$ transition:\vskipone

\noindent $\displaystyle
\Phi_q (z, \zeta, W^2;\mu^2)={1\over2\pi}\int\! dx^-\,
e^{-i z (P^+ x^-)}\,\langle\pi^+(p)\,\pi^-(p')|\,
\bar{\psi}(x^- v')\gamma^+\psi(0)\,|0\rangle.
$\vskip 1cm

$$\epsfxsize 0.7\textwidth\epsfbox{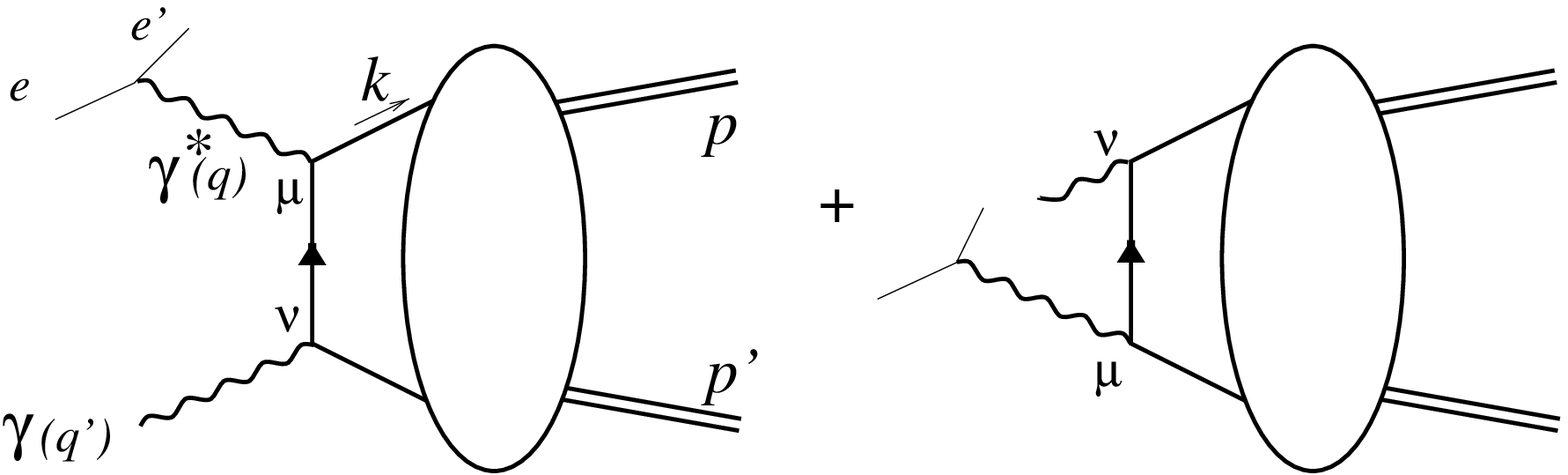}$$
\vskip .7\baselineskip

\noindent Figure~1. $\gamma^*(q)+\gamma(q')\to\pi(p)+\pi(p')$ at
leading order in $\alpha_S$.\vskip 1cm

We find that the $\gamma^\ast \gamma$ amplitude is independent of
$Q^2$ at fixed $\zeta$ and $W^2$, up to small logarithmic scaling
violation. The leading logarithmic corrections to scaling come from
the evolution of the GDA with the factorization scale $\mu^2$. The
framework to study this evolution is the same as the one for standard
distribution amplitudes~[2]. As in the case of quark densities one can
separate $\Phi_q$ in singlet and non singlet components. The singlet
component mixes under evolution with the gluon GDA $\Phi_g$,
associated with the transition $g g \to \pi\pi$. This is further
discussed in~[3].

The process $\gamma^*\gamma\to\pi\pi$ can be considered as the
extension of the pion transition form factor, $\gamma^*\gamma\to
\pi^0$, to a two-pion final state; it may also be seen as the crossed
channel of deeply virtual Compton scattering on a pion target
$\gamma^*+\pi\to \gamma+\pi$.

In electroproduction, $e\gamma \to e \pi\pi$, our process
$\gamma^\ast\gamma\to\pi\pi$ interferes with brems\-strahlung,
$e\gamma\to e\gamma^* \to e\pi\pi$. The latter produces $\pi\pi$ in
the $C$-odd channel, therefore it does not contribute to $\pi^0\pi^0$
production. For a charged pion pair it can be computed from the value
of the timelike elastic form factor measured in $e^+e^-\to\pi^+\pi^-$.
\vskiptwo

\noindent{\bf 3. THE PROCESS $\gamma^*N\to\pi\pi N$}
\vskipone

At large photon virtuality $Q^2$ and small momentum transfer to the
proton the amplitude for electroproduction of a single meson ($\pi,
\rho$, \dots) factorizes~[4] into a hard scattering kernel $H_{ij}$
and two nonperturbative objects obeying their own QCD evolution
equations, namely a nondiagonal parton distribution~[4,5] $f_{i/p}
(x_1,x_1-x,t;\mu^2)$ and the distribution amplitude of the produced
meson $\varphi_j(z;\mu^2)$:
\vskipone

\noindent $\displaystyle {\cal M} =
\sum_{ij}\int_0^1 dz \int_0^1 dx_1\, f_{i/p} (x_1,x_1-x_B,t;\mu^2)\,
H_{ij}(Q^2,x_1/x_B,z;\mu^2)\,\varphi_j(z;\mu^2).$
\vskipone

The introduction of generalized distribution amplitudes allows to
enlarge the scope of this dynamical description to electroproduction
of a nonresonant, low invariant mass $\pi\pi$ pair~[6]. The
factorization diagram is shown in Figure~2. This means for instance
that nonresonant $\pi\pi$ production (and in particular the
$\pi^0\pi^0$ channel absent in $\rho$ decay) should have the same
scaling behavior in $Q^2$ as $\rho$-production.
\vskip 1cm

$$\epsfxsize 0.4\textwidth\epsfbox{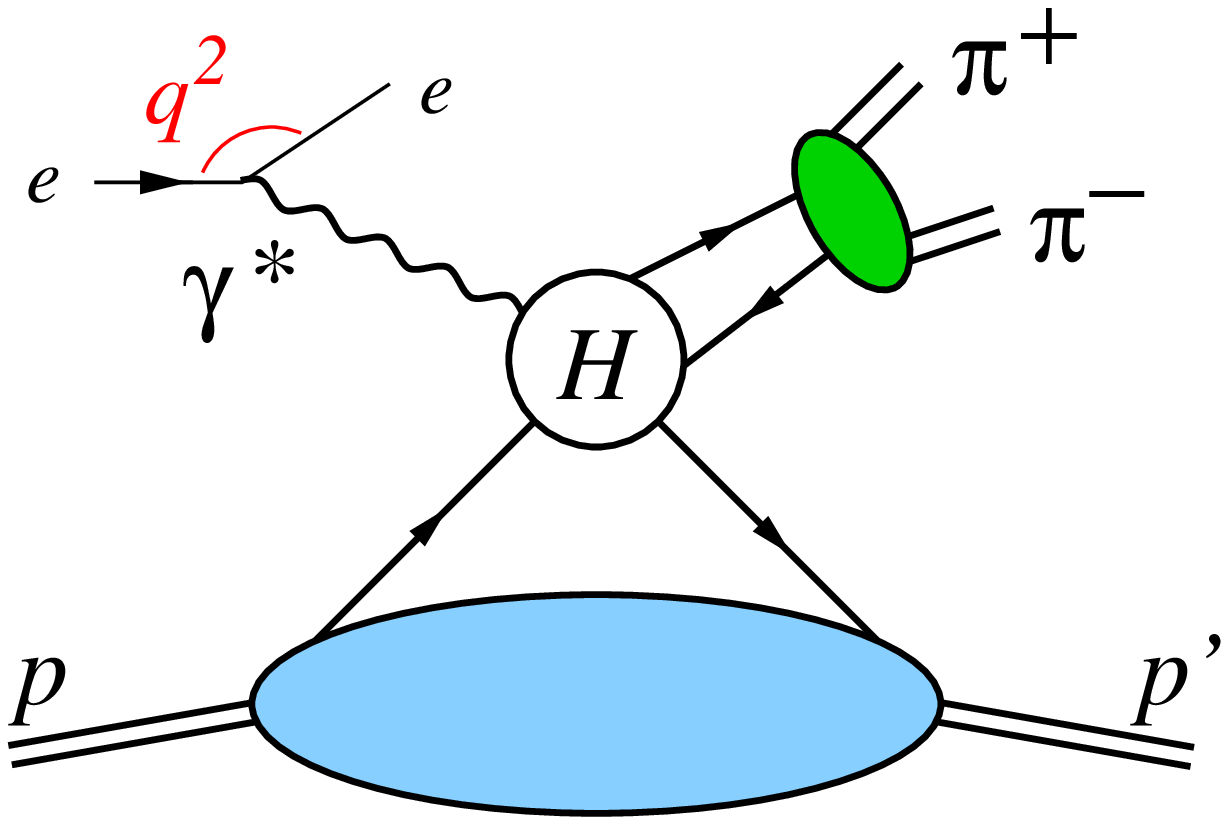}$$
\vskip .7\baselineskip

\noindent Figure~2. Factorization of $\gamma^*+p\to\pi\pi+p$ at large
$Q^2$, small momentum transfer to the proton and small $\pi\pi$
invariant mass.
\vskip 1cm

\noindent{\bf 4. SUMMARY}\vskipone

The process $\gamma^\ast \gamma\to\pi^+ \pi^-$, $\pi^0 \pi^0$, $\dots$
at large photon virtuality and small c.m.\ energy offers a new
opportunity to study hadron structure, where factorization of long and
short distance dynamics enables us to extract nonperturbative
quantities. It should be experimentally accessible at existing or
planned $e^+ e^-$ or $e\gamma$ facilities.

The same long distance amplitudes also appear in other reactions,
namely in exclusive electroproduction of a meson pair, where it allows
an extension of vector meson production studies. The investigation of
these generalized distribution amplitudes complements our existing
tools for the description of hadrons in QCD.
\vskipone

\noindent{\bf Acknowledgments.} 
SUBATECH est l'unit\'e mixte 6457 de l'Universit\'e de Nantes, de
l'Ecole des Mines de Nantes et de l'IN2P3/CNRS. CPhT est l'unit\'e
mixte 7644 du CNRS.
\vskiptwo

\noindent{\bf REFERENCES}\vskipone

{\parindent 0cm
1. M. Diehl, T. Gousset, B. Pire and O.V. Teryaev, Phys.\ Rev.\ Lett.\
   81 (1998) 1782.

2. G.P. Lepage and S.J. Brodsky, Phys.\ Lett.\ B 87 (1979) 359;
   
\hskip 15pt A.V. Efremov and A.V. Radyushkin, Phys.\ Lett.\ B 94
(1980) 245.

3. M.K. Chase, Nucl.\ Phys.\ B 174 (1980) 109.

4. J.C. Collins, L. Frankfurt and M. Strikman, Phys.\ Rev.\ D 56
   (1997) 2982.

5. For a review and references, see X. Ji, J.\ Phys.\ G 24 (1998)
   1181.

6. M.V. Polyakov, hep-ph/9809483.
\par}
\end{document}